\documentclass[fleqn,usenatbib]{mnras}
\usepackage[utf8]{inputenc}
\usepackage{threeparttable}
\usepackage{amsmath}
\usepackage{graphicx}
\usepackage{epstopdf}
\usepackage{threeparttable}
\usepackage{graphicx, color}
\usepackage{bm}
\usepackage{soul}
\usepackage{amssymb}
\usepackage{upgreek}
\usepackage{hyperref}
\usepackage{ulem}

\usepackage{dcolumn}

\newcommand{\eqb}{\begin{eqnarray}}
\newcommand{\eqe}{\end{eqnarray}}

\graphicspath{{figures/}} 

\title[QPOs in Kinked Jets]  
{Kink Instabilities In Relativistic Jets Can Drive Quasi-Periodic Radiation Signatures}

\author[L. Dong, H. Zhang and D. Giannios]{Lingyi Dong$^1$, Haocheng Zhang$^{1}$, Dimitrios Giannios$^1$\\
$^{1}$ Department of Physics, Purdue University, 525 Northwestern Avenue, West Lafayette, IN 47907, USA
}

\begin{document}

\date{Received.../Accepted...}

\pagerange{\pageref{firstpage}--\pageref{lastpage}} \pubyear{2018}

\maketitle

\label{firstpage}

\begin{abstract}
Relativistic jets are highly collimated plasma outflows emerging from accreting black holes. They are launched with a significant amount of magnetic energy, which can be dissipated to accelerate nonthermal particles and give rise to electromagnetic radiation at larger scales. Kink instabilities can be an efficient mechanism to trigger dissipation of jet magnetic energy. While previous works have studied the conditions required for the growth of kink instabilities in relativistic jets, the radiation signatures of these instabilities have not been investigated in detail. In this paper, we aim to self-consistently study radiation and polarization signatures from kink instabilities in relativistic jets. We combine large-scale relativistic magnetohydrodynamic (RMHD) simulations with polarized radiation transfer of a magnetized jet, which emerges from the central engine and propagates through the surrounding medium. We observe that a localized region at the central spine of the jet exhibits the strongest kink instabilities, which we identify as the jet emission region. Very interestingly, we find quasi-periodic oscillation (QPO) signatures in the light curve from the emission region. Additionally, the polarization degree appears to be anti-correlated to flares in the light curves. Our analyses show that these QPO signatures are intrinsically driven by kink instabilities, where the period of the QPOs is associated to the kink growth time scale. The latter corresponds to weeks to months QPOs in blazars. The polarization signatures offer unique diagnostics for QPOs driven by kink instabilities.
\end{abstract}


\section{Introduction}\label{sec:intro}

Relativistic jets are frequently observed in high-energy astrophysical systems, including active galactic nuclei (AGNs). These plasma jets are launched by fast accreting supermassive central black holes in AGNs. Some AGN jets happen to point very close to our line of sight. Such AGN jets, referred to as blazars, appear very bright due to relativistic beaming effects. They are characterized by a very rich phenomenology, which has been actively studied at different wavelengths for decades. 

Blazars show highly variable nonthermal-dominated emission from radio up to TeV $\gamma$-rays, with variability time scales ranging from minutes \citep[e.g.,][]{Ahnen17, Ackermann16} to years \citep[e.g.,][]{Villforth10,Acciari14}. The short variability time scales suggest that the blazar emission comes from compact, sub-parsec regions in the jet, often referred to as the blazar zone, where substantial energy dissipation and particle acceleration take place. Additionally, the radio to optical blazar emission is known to be polarized, with polarization degree (PD) up to $50\%$ \citep{Scarpa97,Smith17}. This is consistent with synchrotron emission by nonthermal electrons in a partially ordered magnetic field \citep{Lyutikov05,Zhang15}. The X-ray to $\gamma$-ray blazar emission is likely from the inverse Compton scattering of the same electrons in the emission region \citep{Maraschi92,Dermer92,Sikora94}, but the recent neutrino detection also hints to hadronic contributions \citep{Icecube18}. Optical polarization monitoring programs have revealed that polarization degree and angle can both be highly variable during blazar flares, such as large polarization angle (PA) rotations \citep{Marscher10,Itoh16,Blinov18}. These observations provide strong evidence that the magnetic field is actively evolving during blazar flares. Very interestingly, some observations have found quasi-periodic oscillation (QPO) signatures for several blazars \citep{Hayashida98, King13, Graham15,Ackermann_2015}. These QPO signatures can happen in various observational bands, which are often considered as evidence for quasi-periodic physical processes in the central engine, such as binary supermassive black holes and periodic changes in the accretion flow.

Blazar jets are likely to be launched with a significant amount of magnetic energy at the central black hole \citep{Blandford_1977, Komissarov_2001,Tchekhovskoy10}. During the jet propagation, the magnetic energy initially stored in the jet can convert into other forms, including the bulk kinetic energy, thermal energy, and nonthermal particle acceleration. However, how exactly the jet magnetic energy evolves during propagation is not well understood. In light of the very bright and highly variable emission from the blazar zone, it clearly marks a special position in the jet propagation and evolution. It is often proposed that blazar flares are driven by shocks \citep[e.g.,][]{marscher85,Boettcher10}. In order for shocks to be efficient in accelerating particles, the blazar zone should be weakly magnetized \citep{Sironi15}. In such kinetic-driven scenarios, the initial high magnetic energy in the jet quickly converts to the bulk kinetic energy, thus the jet itself becomes weakly magnetized before it shines in the blazar zone \citep[e.g.,][]{Achterberg_2001, Nishikawa_2003}. However, shock models have difficulty in explaining the fast variability observed in blazars \citep{Sironi15}. Furthermore, recent simulations suggest that shocks can result in unreasonably high optical PD due to strong compression of the magnetic field at the shock front in a weakly magnetized environment \citep{Zhang16}. Alternatively, the jet may reach the dissipation zone strongly magnetized. In this case, the particle acceleration in the blazar zone can originate from magnetic energy dissipation due to kink instabilities and/or magnetic reconnection \citep{Giannios09,Oneill12,Barniol_Duran17,Giannios19}. Recent particle-in-cell (PIC) simulations have demonstrated that kink instabilities and magnetic reconnection can produce power-law distributions of nonthermal particles, which are consistent with observations \citep{Sironi14,Guo14,Alves18}.

Kink instabilities are a kind of current-driven plasma instability. It causes transverse displacements of plasma and twists the magnetic field structure. Earlier works have shown that in a relativistic jet pervaded by helical magnetic fields, kink instabilities can dissipate significant amount of magnetic energy \citep{Mizuno_2009,Oneill12}. The dissipated magnetic energy then may lead to nonthermal particle acceleration. Moreover, the non-axisymmetric nature of kink instabilities has been studied in magnetic-driven jets with large-scale 3D RMHD simulations \citep[e.g.,][]{Guan14, Porth15, Bromberg_2016, Barniol_Duran17}. Interestingly, they only disrupt the central spine of the jet without affecting the global jet structure and propagation direction. This is very likely due to the returning poloidal magnetic field component that envelops the jet, which can stabilize the MHD instabilities towards the jet boundary. The strongest kinked region in the central spine of the jet thus is a promising location for the blazar zone \citep{Giannios_2006, Bromberg_2019}. Nevertheless, the study of radiation and polarization signatures from kink instabilities has not reached the same level of detail. Radiation studies are limited to an isolated segment of jet undergoing kink instabilities \citep[e.g.,][]{Zhang_2017}. While this treatment may be applicable to individual blazar flares, it overlooks long term variability due to the co-evolution of kink instabilities and the large-scale jet structure. Therefore, it remains unclear how radiation and polarization signatures may appear on relatively long time scales in a jet with kink instabilities, and what physical processes we can tell from observational features. 

In this paper, we make the first attempt to study radiation signatures from kink instabilities in a magnetic-driven jet scenario. We adopt 3D large-scale RMHD simulations from \citet{Barniol_Duran17}, in which a magnetized jet that propagates from the central engine into the surrounding medium. During this process, kink instabilities naturally grow up {in the region where the density profile of the surrounding gas transitions from a steep into a shallow one}. We then conduct comprehensive polarized radiation transfer simulations based on RMHD results to study time-dependent radiation and polarization signatures. Our results show that the jet emission is dominated by a localized region with strongest kink instabilities, which may be identified with the blazar zone in AGN jets. Surprisingly, we discover QPO signatures from this emission region. The QPO signatures are attributed to the quasi-periodic energy release of kink instabilities, and their period is associated to the kink growth time. In Sect.~\ref{sec:setup}, we describe our simulation setup and related physical assumptions. In section Sect.~\ref{sec:QPO_sign}, we present our results on radiation signatures and the physical origin of QPO signatures. 
Finally in Sect.~\ref{sec:discuss}, we summarize our results and discuss implications on observations.

\section{Simulation Setup}\label{sec:setup}

We take a numerical approach to study the radiation and polarization signatures from kink instabilities in a magnetic-driven jet. Our method consists of two components: an RMHD simulation to evolve the jet evolution and a radiation transfer simulation to evaluate observable signatures. The RMHD part is adopted from \citet{Barniol_Duran17}. Since the evolution of kink instabilities is affected by the large-scale jet dynamics, we simulate the jet propagation from the central engine into the surrounding medium, where kink instabilities naturally grow up. In this way, we can go beyond the simple one-zone emission model and better understand radiation and polarization signatures from kink instabilities in a more realistic setup.

\subsection{RMHD Setup}\label{sec:MHDsetup}

We adopt the simulation labeled as A2-3D-hr in \citet{Barniol_Duran17}. This run shows quick development of kink instabilities as the jet propagates through surrounding medium and has best spatial resolution compared to other runs in that paper. In the following we discuss briefly its setup. The simplified central engine is assumed to be a perfectly conducting sphere of radius $r_0=l_0$, where $l_0$ is the code length unit, threaded by radial magnetic field lines. The jet is highly magnetized at its base, where the initial magnetization factor $\sigma$, which is the magnetic energy density over enthalpy, is $\sigma_0 = 25$. The sphere is rotating at a constant angular frequency of $\Omega=0.8c/r_0$. This is to mimic a supermassive black hole and its accretion disk rotation. The rotation then coils up the radial magnetic field lines into helices and launch two oppositely directed magnetized jets into the surrounding medium. Initially, the surrounding medium is cold and static. In the neighborhood of the central engine, the surrounding medium is expected to be dominated by winds from the accretion disk, resulting in a steep density profile, but at larger distances the interaction between jet/wind and the interstellar medium can become important resulting in a possibly flatter density profile. This motivates our assumed broken power-law density profile for the gas, where the break point is at radius $r_B=100l_0$, so that 
\begin{equation}
\rho=
\left\{
\begin{array}{lcc}
\rho_0(\frac{r}{r_0})^{-3} & , & r\leq r_B \\
\rho_0(\frac{r_B}{r_0})^{-3}(\frac{r}{r_B})^{-1} & , & r> r_B.
\end{array}
\right.
\end{equation}
As demonstrated in \citet{Barniol_Duran17}, the jet recollimates at the density break (i.e., around the transition between the steep and flat density profiles) and consequentially becomes narrower at the recollimation point. Such site is natural for kink instabilities to develop since it is where the jet's cross section is reduced. Indeed, \citet{Barniol_Duran17} successfully stimulated kink instabilities in the jet using this density break. Although kink instabilities may naturally grow in a magnetized jet with various initial setup parameters and boundary conditions \citep[e.g., ][]{Guan14}, we choose to adopt the simulation in \citet{Barniol_Duran17} mainly because it produces adequately resolved fully grown kinked jet within a reasonable amount of computational resources.

The time-dependent 3D RMHD simulation is performed with the HARM code \citep{Barniol_Duran17}. It takes a non-uniform spherical grid of $512\times 192\times 384$ in $r$, $\theta$, and $\phi$ directions, where the $r$ direction is in logarithmic scale and the $\theta$ grids are more concentrated in the jet propagation direction. The simulation is open boundary with a large box size $r_{max}=10^5l_0$ to avoid any outer boundary effects on the jet.

\begin{figure}
\centering
\includegraphics[width=0.45\textwidth]{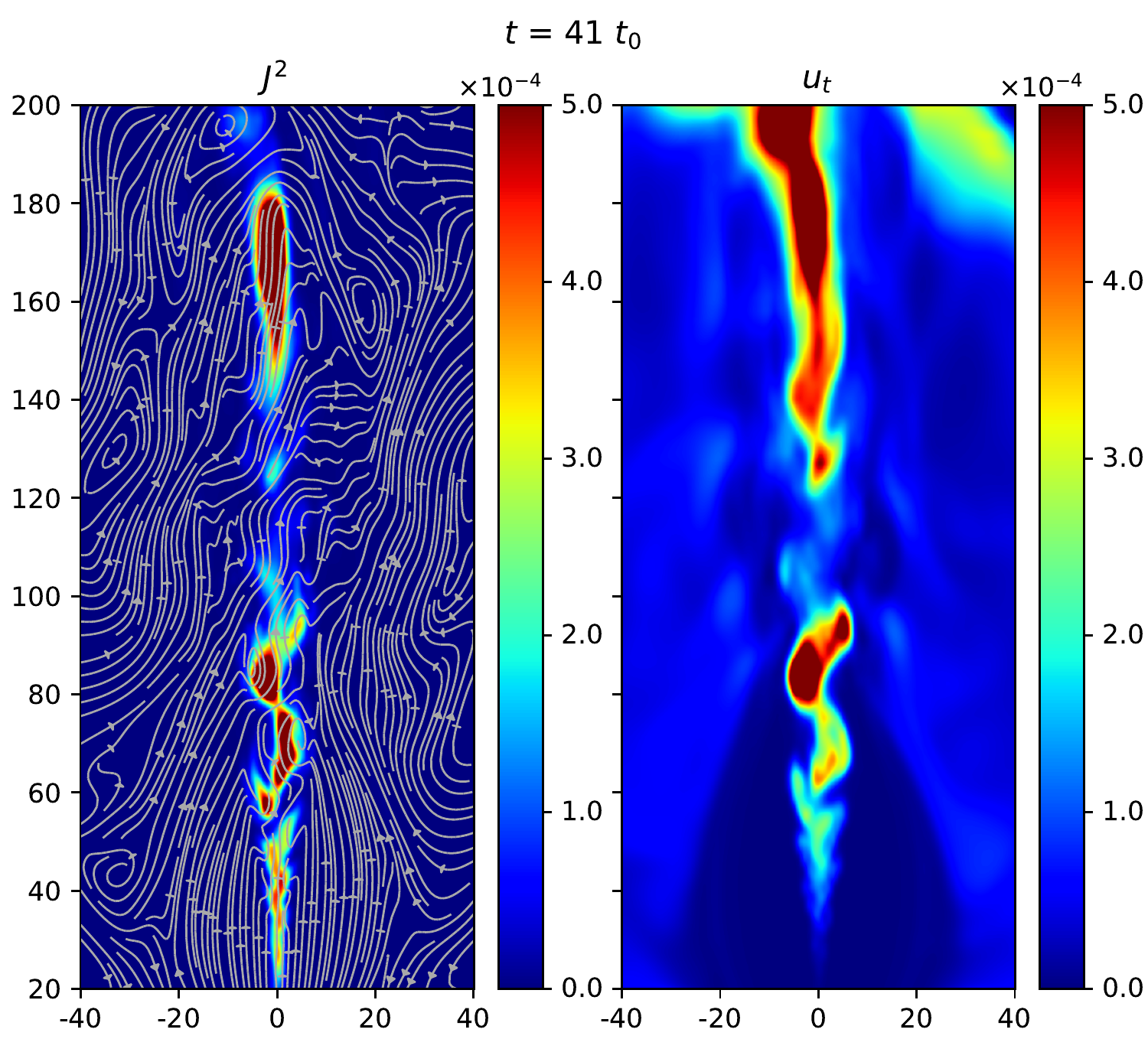}
\caption{A snapshot of the jet at $t = 41 t_0$ from $r=20l_0$ to $r=200 l_0$ in jet propagation direction, with a radius of $40 l_0$. At this snapshot, the jet head has mostly moved out of $r=200 l_0$. \textbf{Left Panel: } the current density \textbf{with magnetic fields lines projected on the same plane}. The twisted structures indicate the location of kink instabilities. \textbf{Right Panel: } thermal energy density.}
\label{fig:jet}
\end{figure}

\subsection{Radiation Transfer Setup}\label{sec:RTsetup}

For typical blazar zone parameters, the infrared to optical blazar emission, which show rich variability patterns, are in the optically thin regime. Therefore, considering only synchrotron emission with frequencies beyond near-infrared, our calculation ignores the synchrotron-self absorption and Faraday rotation effects. In this situation, the polarization signatures directly reflects the magnetic field evolution in the emission region. To calculate the synchrotron emission, we need to know both the magnetic field and nonthermal particle distributions in each simulation cell. While the former is directly given by RMHD simulations, the latter requires additional modeling. Recent PIC simulations have shown that kink instabilities can accelerate nonthermal particles via unscreened electric fields that develop in the unstable regions \citep{Alves18}. Additionally, it has been suggested that kink instabilities can considerably twist the magnetic field lines, which may generate current sheets and trigger magnetic reconnection \citep{Begelman98,Giannios_2006}. Magnetic reconnection then accelerates particles into power-law distributions \citep{Sironi14,Guo14}. In our setup, the acceleration of nonthermal particles can, therefore, originate in regions of dissipation of jet magnetic energy. In our ideal RMHD simulation, the thermal energy is a good indicator of the location of magnetic energy dissipation, so that the nonthermal particle energy can be considered to be a fraction of the thermal energy released in the jet due to kink instabilities. Indeed, as we can see in Figure \ref{fig:jet}, the kinked region shows significant thermal energy (see details in the next section). Therefore, we take the local thermal energy density as a normalization of the nonthermal particle energy density in each simulation cell. 

Kink instabilities evolve on the time scale of Alf\'ven wave crossing time of the cross section of the jet, which is comparable to the light crossing time scale in a considerably magnetized environment. Therefore, we target at relatively long-term variability signatures. In this case, the detailed nonthermal particle acceleration and radiative cooling processes are of much shorter time scales. Take the blazar jet as an illustrative example, the light crossing time scale of the blazar emission region in the comoving frame is typically on the order of $10^6-10^7~\rm{s}$, based on typically observed days to weeks variability of blazars. However, the nonthermal electrons in a typical leptonic jet model with a magnetic field strength of $0.1- 1 \rm G$ in the blazar zone has synchrotron cooling time scales of $\sim 10^5~\rm{s}$ for emission in optical bands. In light of the large difference in time scales, the nonthermal particle evolution may not play an important role on long-term variability. We thus take the simplification that the nonthermal particles everywhere in the simulation box are of the same power-law distribution with an exponential cutoff,
\begin{equation}
n(\gamma)=n_0\gamma^{-2}e^{-\frac{\gamma}{5\times 10^4}}~~.
\end{equation}
in which $n_0$ is a normalization factor chosen by
\begin{equation}
\int^{\infty}_1 n(\gamma)d\gamma =0.05u_t~~,
\end{equation}
so that we normalize nonthermal particles based on the local thermal energy density $u_t$. Additionally, we take an output snapshot of the RMHD simulation every $t_0=20l_0/c$ (about $10^6~\rm{s}$ for typical blazar parameters), which is significantly longer than the nonthermal particle acceleration and cooling time scales. At every RMHD output, the nonthermal electrons will be refreshed based on the new physical parameters in each simulation cell.

Given the magnetic field and nonthermal particles, we carry out time-dependent polarized radiation transfer using the 3DPol code \citep{Zhang14}. This code evaluates the polarization-dependent synchrotron emission and performs radiation transfer via ray-tracing method, which naturally includes light crossing time effects. This code considers time-, space-, and frequency-dependencies of the synchrotron emission. We also make use of its recent upgrade with polarized emission map on the plane of sky \citep{Zhang18}, which can help us to better analyze the radiation signatures from kink instabilities. Since 3DPol uses a uniform Cartesian coordinates for best performance of the polarized radiation transfer, we interpolate the non-uniform RMHD grids to uniform Cartesian grids with a resolution of $0.8l_0$. We find this resolution sufficient to capture all the details of fluid dynamics that may affect the radiation signatures.

Due to limited computational resources, \cite{Barniol_Duran17} do not trace the full evolution of the jet from the central engine out to scales several (4-6) orders of magnitude larger where a highly relativistic jet has developed. In this selected RMHD simulation run, the strongest kinked region is located at a length scale that is a factor 100 larger than the injection scale and is characterized by a modest average Lorentz factor $\sim 2$, while typically blazar jets have bulk Lorentz factor $\Gamma \sim 10$ or beyond.
In the following, all light curves, polarization signatures, and snapshots of the polarized emission maps on the plane of sky are presented in the comoving frame of the jet for better illustration purposes.
The viewing angle is set to be $90^{\circ}$ from the jet propagation direction in the comoving frame by default. We define the PA in the following way. PA $=0$ if the electric field vector of the observed electromagnetic wave preferentially oscillates along the jet projection on the plane of sky. PA $=90^{\circ}$ is then in the perpendicular direction. Notice that the PA has $180^{\circ}$ ambiguity. In Section~4, we discuss how these results translate into observer's frame quantities for realistic relativistic boost of $\Gamma \sim 10$ expected in blazars.

We summarize our key physical assumptions for our simulation setups as follows:
\begin{enumerate}
    \item The jet launches from a simplified central engine of a rotating sphere threaded with magnetic field lines.
    \item The surrounding density follows a prescribed broken power-law profile.
    \item The initial jet plasma and the surrounding medium are cold and laminar.
    \item The nonthermal particles follow a power-law distribution whose energy density is proportional to the local jet thermal energy density. They are refreshed at each RMHD output time step.
\end{enumerate}

\section{QPO Signatures Arising From Kink Instabilities}\label{sec:QPO_sign}

In this section, we present our simulation results of radiation and polarization signatures in the jet emission region, based on global RMHD simulation of a magnetized relativistic jet with kink instabilities. We find that most of the emission comes from a localized region in the jet with strongest kink instabilities. We identify this region as the blazar emission region. Very interestingly, we find QPO signatures in radiation signatures. In the following, we will analyze the physical processes behind the radiation patterns and investigate the origin of QPO signatures.

\subsection{Jet Evolution and Temporal Radiation Signatures}

The jet simulation contains a magnetized, rotating spherical central engine at the inner boundary. The rotation coils up the magnetic field lines attached to the central engine, and launches relativistic jets with high magnetic energy, which then penetrate through the surrounding medium. The jet expansion is fairly quick. The jet head reaches $r\gtrsim 200l_0$ at $t\sim 35t_0$, so that the jet is adequately large to study its structure and radiation signatures. Figure~\ref{fig:jet} plots a vertical cut of the RMHD jet simulation at a more mature stage ($t=41t_0$). At this snapshot, the jet head has mostly moved out of $r=200l_0$. We can easily identify kink instability signatures in the current density plot on the left panel. As illustrated by the magnetic field streamlines, kink instabilities disrupt the central spine of the jet, implying dissipation of magnetic energy. As a result, we observe concentration of thermal energy on the right panel that traces the kinked current density, which marks the location of magnetic energy dissipation. Interestingly, the strongest kinked region, which also exhibits the highest thermal energy density, concentrates between $r\gtrsim 60l_0$ and $r\lesssim 100l_0$. We notice that another region with concentrated thermal energy shows up above $r\gtrsim 120l_0$, but there is no clear signature in the current density plot for kink instabilities. We attribute this heated region to the passage of the jet head, which may cause shocks that heat up the plasma. This is evident by the right panel of Figure~\ref{fig:jet}, where we can still see some of the enhanced thermal energy outside the central spine of the jet between $175l_0$ and $200l_0$.

\begin{figure*}
    \centering
    \includegraphics[width=0.49\textwidth]{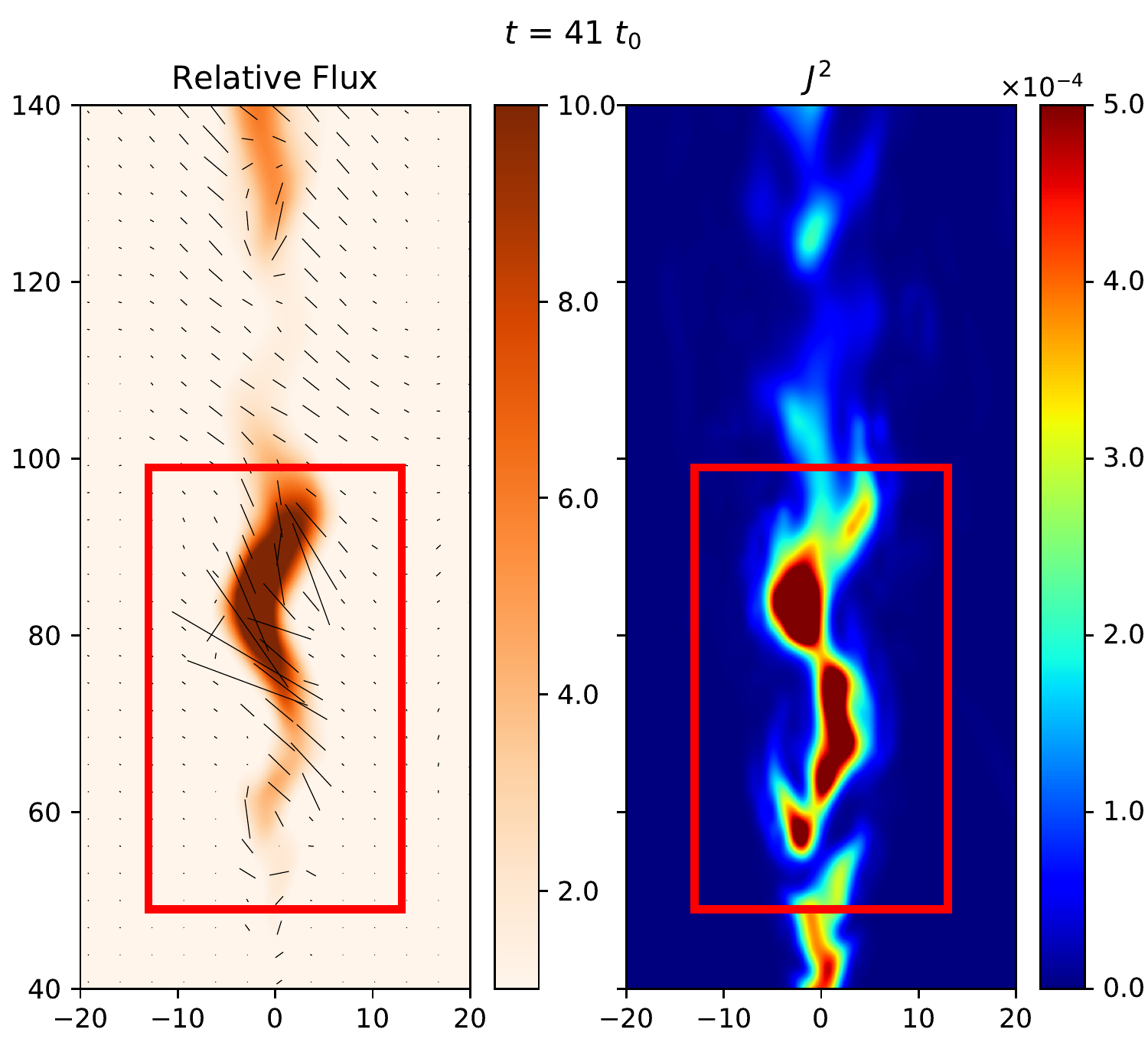}
    \includegraphics[width=0.49\textwidth]{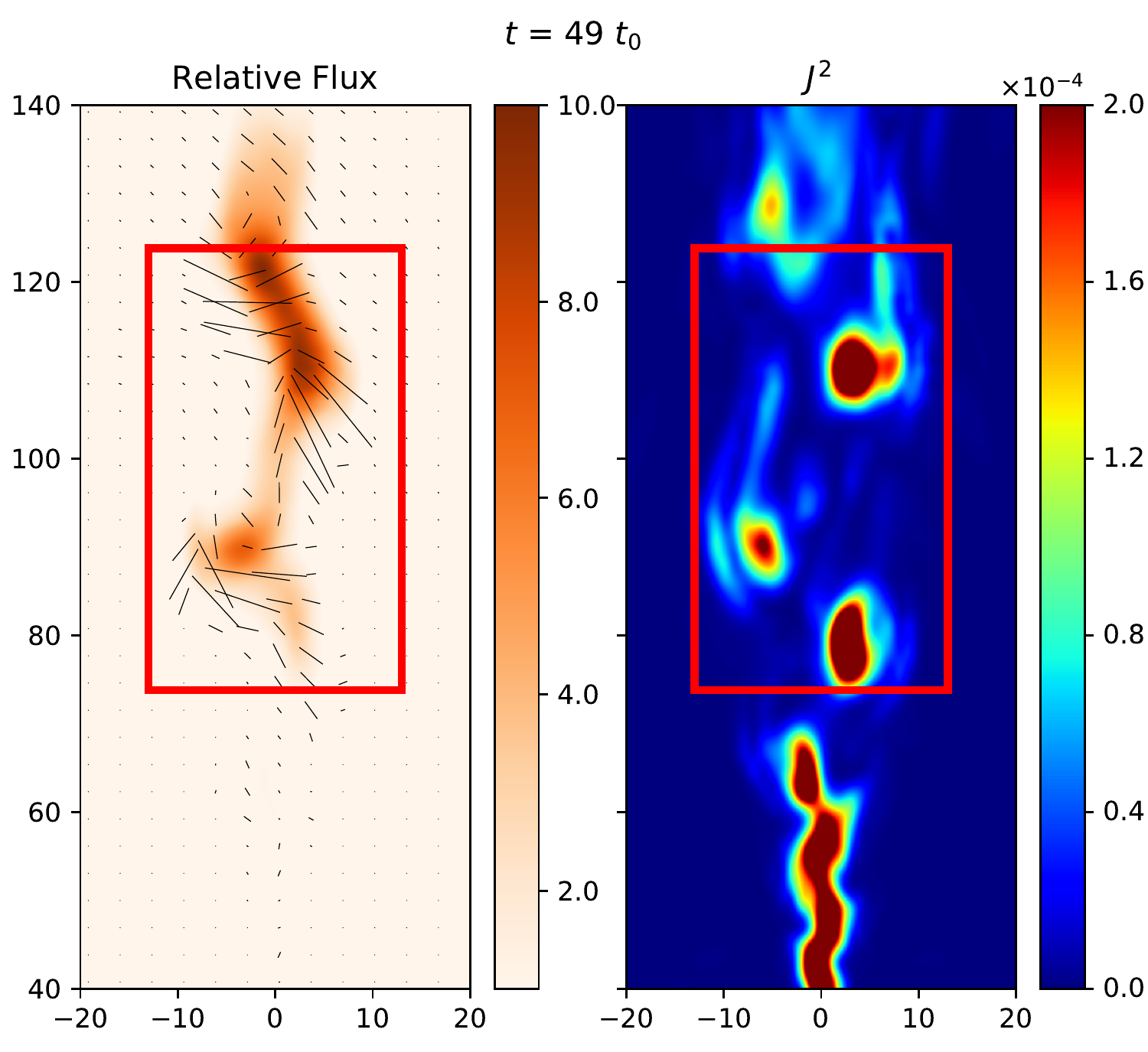}
    \caption{Left Panel: Snapshots of the jet with polarized emission map (left) and current density (right) at time step $t = 41t_0$. The red tracing box highlights the emission region. Right panel: Same plots of the jet at time step $t = 49 t_0$. The emission region/tracing box moves $\sim 25 l_0$ downstream the jet. The length of overplotted black bars represent local PD, and their orientations represent corresponding local PA.}
    \label{fig:box_move}
\end{figure*}

We limit our radiation transfer simulation between $r=20l_0$ and $r=200l_0$. This is because below $r=20l_0$, the injection at the central engine can strongly affect the physical properties of its neighborhood. Beyond $r=200l_0$, since we use logarithmic grids in $r$, the resolution of RMHD simulation is not adequately good to derive radiation and polarization signatures. Figure~\ref{fig:box_move} shows two snapshots of the polarized emission map and current density of the jet. It can be seen that most of the emission concentrates at the strongest kinked region, which appear similar to the kinked current density distribution. Additionally, the polarization vectors indicate that the underlying magnetic field morphology also follows the kinked structure. These features are consistent with previous studies on kink instabilities \citep{Zhang_2017,Nalewajko17}. We notice that the emission region appears to slowly move upward. It is because in our simulation setup, the central engine continuously pumps energy into the jet, so that the jet keeps expanding and pushing external medium outward. This is evident by the expansion of the emission region in Figure~\ref{fig:box_move}, which shows that the emission region not only moves upward but also expands laterally. The location and expansion of the emission region also illustrate that although we include a density break in the external density profile to stimulate the kink growth, the subsequent kink evolution and the location of the strongest kinked region is not determined by the location of the initial density break but by the subsequent jet-medium interactions.

\begin{figure}
\includegraphics[width=0.45\textwidth]{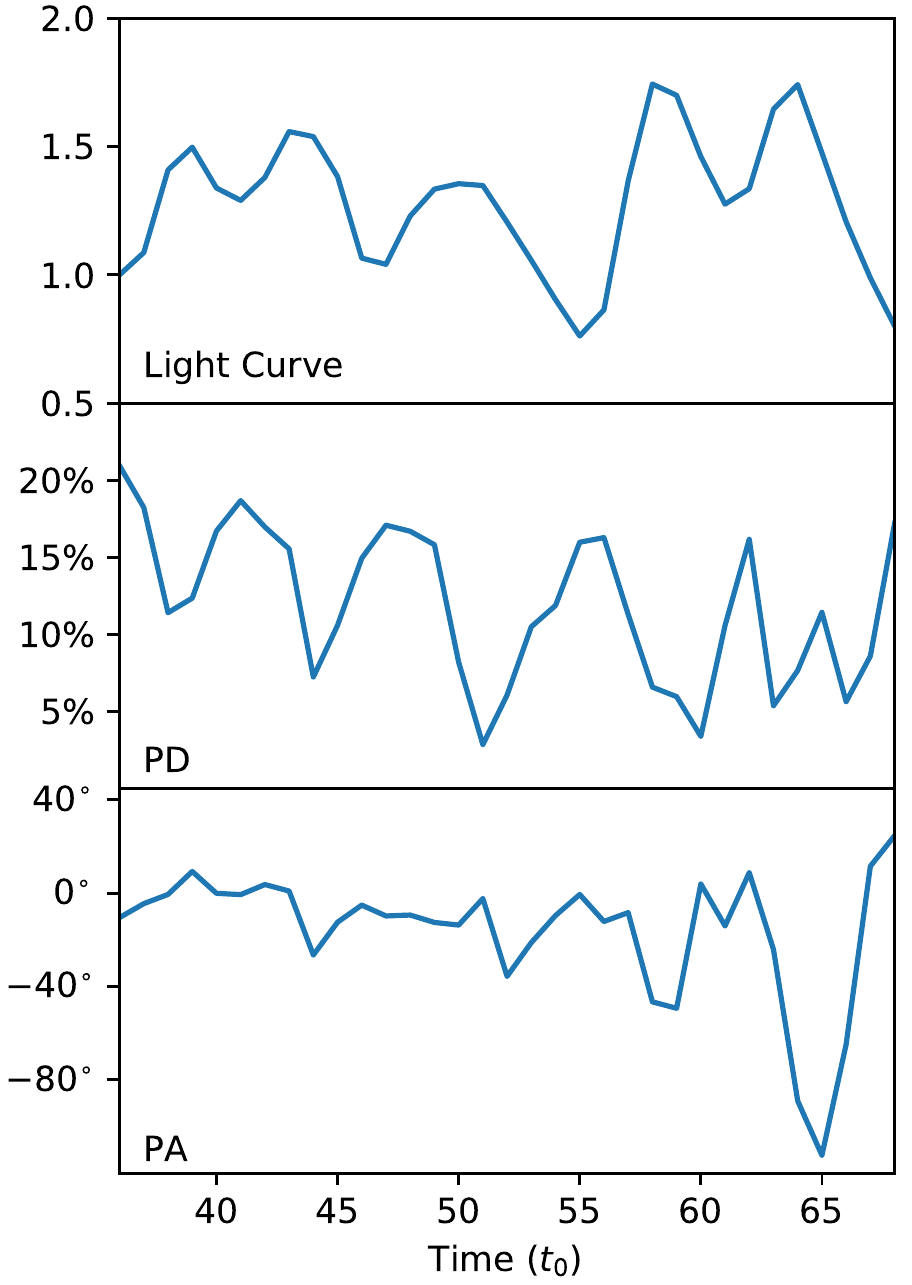}
\caption{Observed QPO features in light curve (top panel), PD (middle panel) and PA (bottom panel). The period of the light curve and PD are close to $\sim 6 - 7 t_0$.}
\label{fig:qpo}
\end{figure}

Given that the emission region is not of a regular shape and can slowly move radially outwards, we evaluate the time-dependent radiation signatures in the emission region by tracing it in a moving box, highlighted in red in Figure~\ref{fig:box_move}. Figure~\ref{fig:qpo} shows the resulting light curve, PD and PA of the jet synchrotron radiation in the optically thin regime.  We observe five flares in the light curve. During the flares, the flux amplitude approximately doubles, indicating significant energy release due to kink instabilities. Most importantly, these flares have similar duration and gaps between flares, appearing to be quasi-periodic (see Section \ref{sec:sub_qpo} for details). The PD averages at $\sim 10\%$ and remains below 20\% throughout the simulation, which is consistent to general blazar optical polarization observations \citep[e.g.,][]{Angelakis16}. Interestingly, the PD appears to be anti-correlated to the light curve. The PA mostly stays at $\sim 0$, but it shows a major jump from $\sim 0$ to $\sim -90^{\circ}$ then back to $\sim 0$ near the end of our simulation. This PA jump happens in coincidence with a small rise in the PD at the same time.

\begin{figure*}
\includegraphics[width=0.45\textwidth]{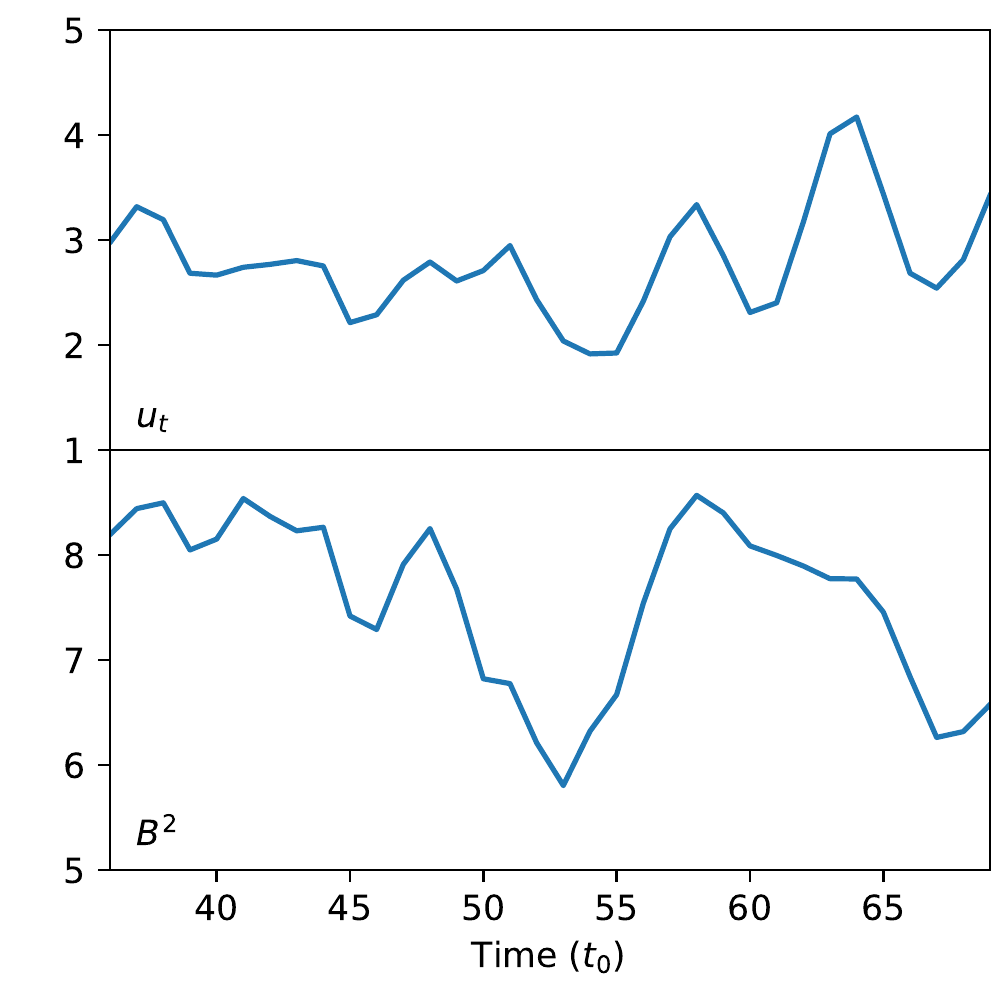}
\includegraphics[width=0.45\textwidth]{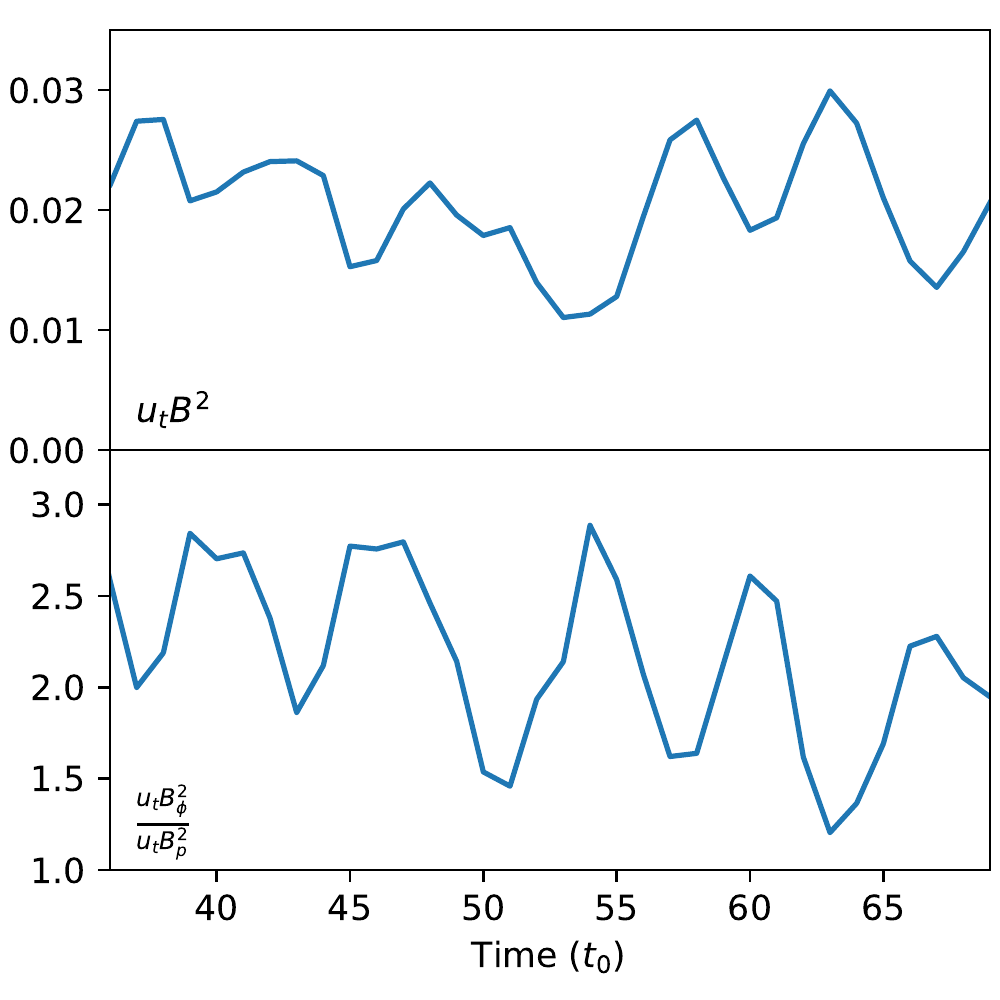}
\caption{
Plots of $u_t, u_t B^2$ and $\frac{\sum{u_t B_y^2}}{\sum{u_t B_x^2}}$ in the tracing box (the emission region). With the Exception of $B^2$, all other curves show quasi-periodic signatures, as discussed in Section~\ref{sec:sub_qpo}. \textbf{All quantities are in code units.}}
\label{fig:qpo_causes}
\end{figure*}

The above polarization variations are due to the magnetic field evolution during kink instabilities. As we can see in Figure~\ref{fig:qpo_causes} (lower right panel), the synchrotron contribution due to the toroidal component is always stronger than the poloidal component. Therefore, the PA is generally aligned with the jet propagation direction, which is set to be 0 in our PA definition. However, at $t\sim 63 t_0$, the ratio between the two contributions gets very close to 1, implying that the toroidal contribution is not so dominating. This leads to a major rotation of the PA towards $-90^{\circ}$. Nevertheless, the poloidal contribution is not adequate enough to dominate over the toroidal contribution, and the ratio quickly restores to a toroidal dominating situation. Therefore, we observe that the PA rotates back to $\sim 0$ without completing a full PA swing. We can also observe in the right two panels in Figure~\ref{fig:qpo_causes} that this ratio between toroidal and poloidal contribution is anti-correlated to $u_tB^2$, which represents the synchrotron emissivity. This suggests that when kink instabilities are efficiently releasing magnetic energy, they strengthen the synchrotron contribution from the poloidal magnetic component. This finding is consistent with \citet{Zhang_2017}, and explains the anti-correlation between PD and light curve.

\subsection{Observed QPO features and causes}\label{sec:sub_qpo}

Both the light curve and PD show QPO signatures in our results. To further examine the QPO patterns, we employ Markov-Chain Monte Carlo (MCMC) method to analyze its period. We obtain a period of $6.5 t_0$ for the light curve and $6.8 t_0$ for the PD with a simple sinusoidal model. The percentage difference in fitted periods is very small ($< 5\%$), thus we take $T=6.5 t_0$ as the period for both light curve and PD. In addition, an anti-correlation between light curve and PD is observed through the phase difference of $\sim \pi$ in their best-fit sinusoidal curves.

We can see in Figure~\ref{fig:qpo_causes} that $u_tB^2$, which represents the instantaneous synchrotron emissivity in the emission region without considering light crossing effects, appear to be quasi-periodic. To further understand what causes this pattern, we plot $u_t$ and $B^2$ in the emission region (Figure~\ref{fig:qpo_causes} left panels). Here to trace the 3D emission region in the jet, we add an extra dimension in depth that is equal to the width of the tracing box on the plane of sky. Apparently, it is the thermal energy density $u_t$ that behaves quasi-periodically. Since the thermal energy in the strongest kinked region comes from the dissipation of magnetic energy, this suggests that the conversion is quasi-periodic. Notice that the curves plotted in Figure~\ref{fig:qpo_causes} are shifted to earlier time in comparison to the light curve and PD. Also the relative flare amplitudes between each peak in $u_tB^2$ and the light curve appear slightly different. This comes from the light crossing time delay. Since the light crossing time of our simulation box is $\sim 2t_0$, it makes all observables delay $\sim 2t_0$ than the instantaneous physical parameters in the emission region. Also as we can see in Figure~\ref{fig:box_move}, the emission region spans about $25l_0$, so that within the emission region it has a light crossing time delay of $\sim t_0$. This leads to a shift the relative flare amplitude between peaks in the light curve compared to $u_tB^2$. On the other hand, the magnetic energy in the tracing box does not clearly exhibit the same quasi-periodic behaviour (Figure~\ref{fig:qpo_causes} lower left). This is because in addition to the quasi-periodic conversion, the magnetic energy inside the tracing box also includes that flow of magnetic energy in and out of the tracing box, which is not in general quasi-periodic.

\begin{figure}
\includegraphics[width=0.45\textwidth]{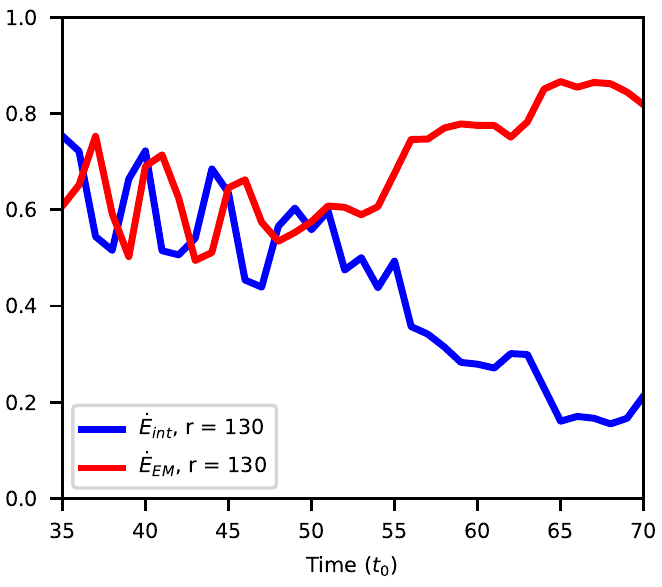}
\caption{Thermal $\dot{E}_{\rm INT}$ and Poynting $\dot{E}_{\rm EM}$ flux versus time at a fixed distance $r = 130 l_0$. Notice out-of-phase oscillations in both $\dot{E}_{\rm INT}$ and $\dot{E}_{\rm EM}$. Such oscillations in these two quantities show the periodic conversion of magnetic energy into thermal energy. At times $t> 50t_0$, the dissipation region moves beyond the $r = 130 l_0$ distance and the oscillations stop.} 
\label{fig:checkpoints}
\end{figure}

To further examine the quasi-periodic magnetic energy conversion to thermal energy, Figure~\ref{fig:checkpoints} plots the internal and electromagnetic energy flux at a fixed distance $r=130l_0$. This location shows clear signatures of kink instabilities after $t=35t_0$. We can clearly see out-of-phase, QPO signatures in both curves. This demonstrates that indeed kink instabilities are converting magnetic energy into thermal energy quasi-periodically.

But what causes the periodic magnetic energy dissipation during kink evolution? Since the kink nodes appear to be a quasi-periodic structures, as shown in many previous works \citep{Mizuno_2009,Oneill12}, We believe that the period of QPOs is associated to the growth time of kink instabilities. \citet{Mizuno_2009} have shown that the kink growth time can be estimated by the evolution of the transverse motion. To quantify this time scale, we estimate it to be the ratio of the transverse displacement of the jet from its central spine over the averaged transverse velocity, i.e., $\tau_{\rm KI} \sim R_{\rm KI}/\langle v_{\rm tr} \rangle$, where $R_{\rm KI}$ is the transverse displacement of the strongest kinked region and $\langle v_{\rm tr} \rangle$ is the average transverse velocity. We find that $R_{\rm KI}$ is about $\sim 20 l_0$ in our simulation and $v_{\rm tr}\sim 0.16c$. 
Therefore, the growth time of kink instabilities is $\tau_{\rm KI} \approx 125 l_0/c$, which is consistent to the QPO period at $T\sim 6.5t_0=130 l_0/c$ in the comoving frame.

\begin{figure}
\includegraphics[width=0.45\textwidth]{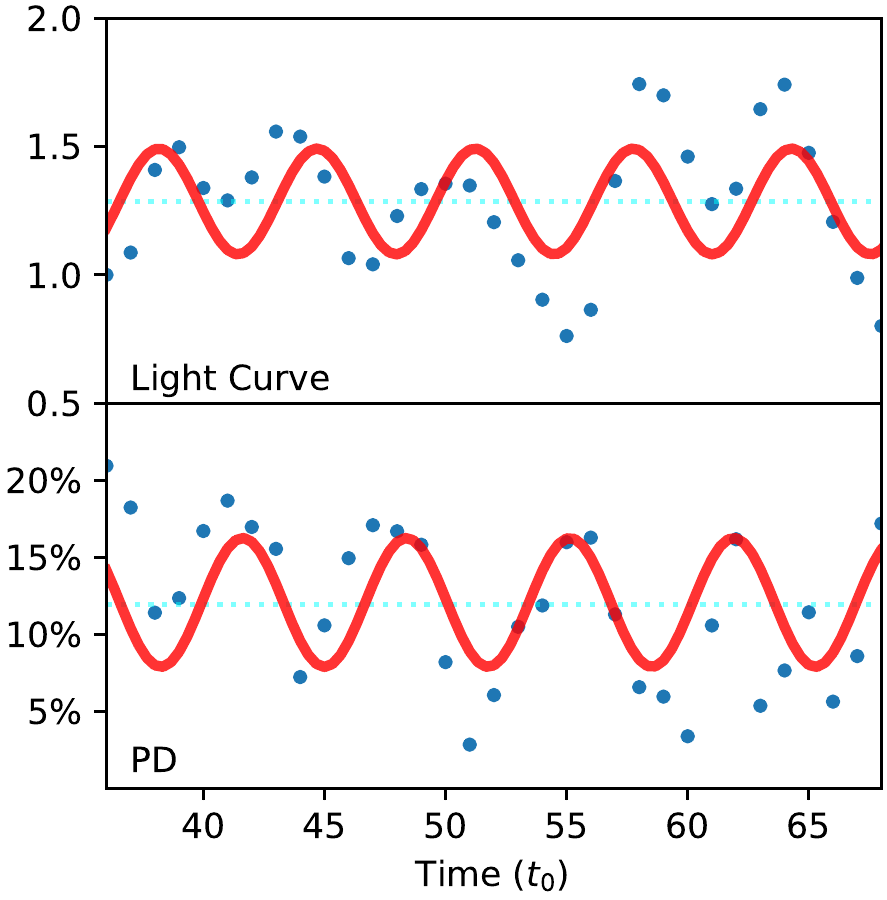}
\caption{Fitting of light curve and PD variations with Markov-Chain Monte Carlo \citep{emcee} with a sinusoidal model.The period of QPO in the light curve is $6.53^{+0.09}_{-0.09} t_0$ and that in PD variation is $6.76^{+0.09}_{-0.09} t_0$. The phase difference of $\sim \pi$ between light curve and PD variations suggests an anti-correlation of the two quantities.}
\label{fig:qpo_mcmcfit}
\end{figure} 

\subsection{Viewing Angle}\label{sec:viewangle}

\begin{figure}
\includegraphics[width=0.45\textwidth]{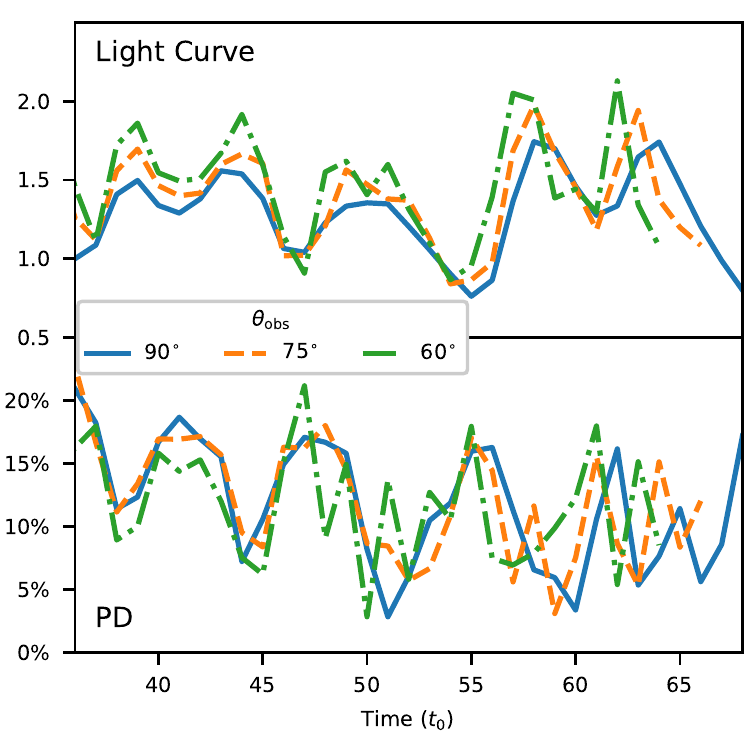}
\caption{Light curves and PD variations with 3 different view angles ($90^{\circ}, 75^{\circ}$ and $60^{\circ}$, with respect to jet propagation in the comoving frame). The QPO signatures persist at different viewing angles.
}
\label{fig:diff_va}
\end{figure}

The above calculation is based on a viewing angle at $90^{\circ}$ from the jet propagation direction in the comoving frame. In principle, the viewing angle can affect the light crossing and projection of the magnetic field on the plane of sky, which can alter radiation signatures. To examine these effects, we consider two more viewing angles, at $75^{\circ}$ and $60^{\circ}$ from the jet propagation direction in the comoving frame. Figure~\ref{fig:diff_va} plots the resulting light curve and PD. PA is not plotted because at both viewing angles it shows small erratic fluctuations around 0.

We observe two main differences. First, the QPO period appears shortened with smaller viewing angles. This is due to the combined effects of the light crossing delay and motion of the emission region. At $\theta_{\rm obs} = 90^{\circ}$, the motion of the emission region is perpendicular to the line of sight, which does not affect observational signatures. However, at smaller $\theta_{\rm obs}$, the upward motion of the emission region has a component along the line of sight, making the emission region closer to the observer. This reduces the light crossing delay, which leads to a shortening of the period by a factor of $1 - \frac{v_{\rm LOS}}{c}$, where $v_{\rm LOS}=v\cos \theta_{\rm obs}$ and $v$ is the speed of the emission region. We remind readers that the motion of the emission region in our simulation is due to the continuous injection of magnetic energy at the base of the jet, which pushes the jet upward. In practice, whether the emission region may move in space can depend on several physical processes, such as the accretion flow variations and its interactions with the ambient gas. Nevertheless, we can clearly see that this motion only affects the period but not the presence of the QPO pattern.

Second, the PD loses QPO behaviors at different viewing angles. This is because the projection of magnetic field onto the plane of sky is altered at different viewing angles. Nonetheless, the PD is still generally anti-correlated with the light curve, where the PD is on average lower at the peak of the flare than the low state. Therefore, polarization signatures can still be powerful diagnostics of whether the QPOs in the light curve is due to kink instabilities. 

\section{Summary and Discussion}\label{sec:discuss}

To summarize this paper, we have found QPO signatures in radiation driven by kink instabilities in relativistic jets, based on our combined large-scale RMHD simulation and polarized radiation transfer. The RMHD simulation self-consistently tracks the jet propagation into the surrounding medium and the development of kink instabilities in the central spine of the jet. It is clear that most of the emission comes from a localized region with the strongest kink instabilities based on our radiation transfer simulation. We find QPO signatures from this emission region. The QPO signatures result from the quasi-periodic dissipation of magnetic energy due to kink instabilities, whose period is associated to the kink growth time. Provided typical blazar jet parameters, we estimate that the typical period of QPOs driven by kink instabilities is on the scale of weeks to months in the observer's frame (see below). In particular, polarization signatures can be powerful diagnostics to identify kink instabilities in jet, where we find anti-correlated PD with the light curve. In some cases, the PD may appear QPO as well, and the PA may exhibit large rotation. In the following, we will discuss the robustness of the QPOs we find with our analyses, and implications on observations.

\subsection{Robustness of QPOs}

Other than the viewing angle, the choice of the tracing box may affect the radiation signatures in our radiation transfer simulation. Due to the irregular shape of the emission region, we have to use a tracing box to evaluate the radiation and polarization signatures from kink instabilities. We cannot sum up all radiation signatures from the simulation box, because a significant amount of emission may originate from the passage of the jet head (see Figure~\ref{fig:box_move}). Additionally, in our simulation the emission region keeps moving up in the jet propagation direction, so that the tracing box has to move accordingly. Thus it is necessary to investigate the robustness of QPOs with different tracing boxes.

\begin{figure}
\includegraphics[width=0.45\textwidth]{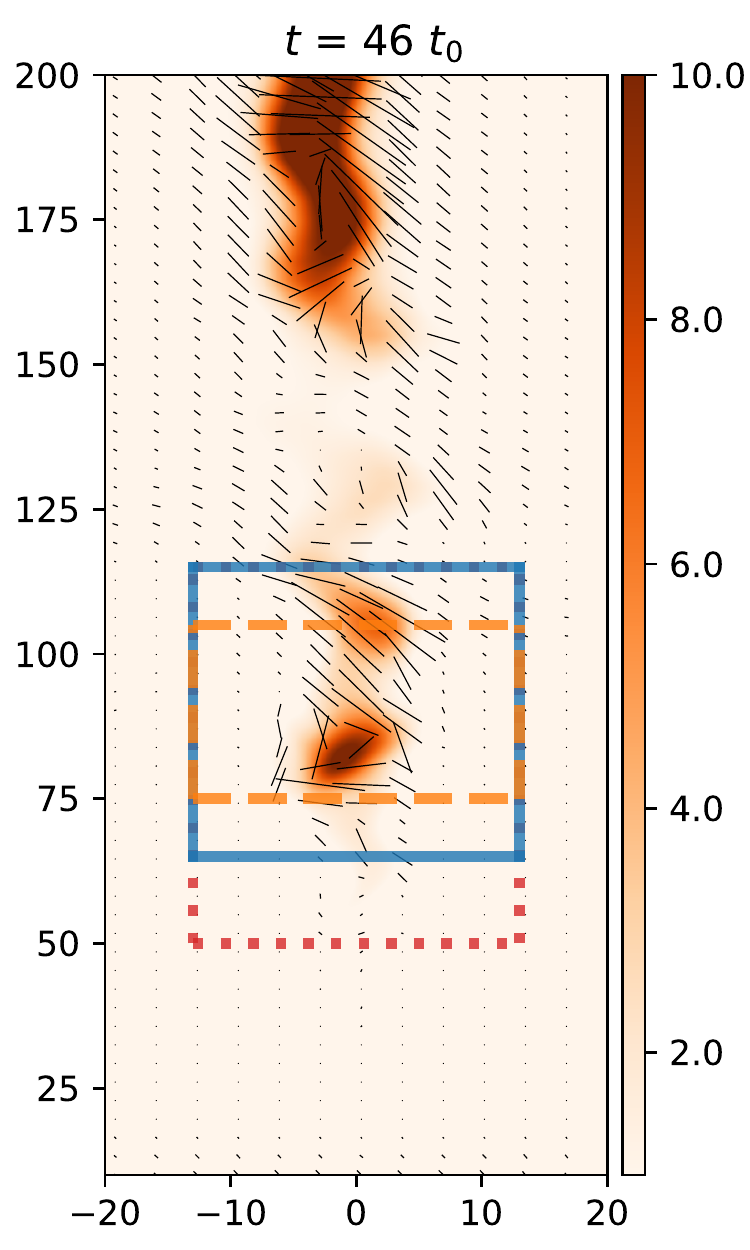}
\caption{Four different treatments of the emission region are shown. \textbf{Solid Blue:} A fixed-middle-size emission region moving downstream the jet at a constant speed. \textbf{Dashed Orange:} A fixed-small-size emission region moving downstream the jet at a constant speed. \textbf{Dotted-Dashed Green:} A stationary fixed-large-size emission region that can span the whole emission region in most time steps. \textbf{Dotted Red:} A emission region that is fixed at the bottom, but its top is moving downstream at a constant speed. This set up attempts to avoid the head of the jet in earlier time steps. }
\label{fig:box_sizes}
\end{figure}

\begin{figure}
\includegraphics[width=0.45\textwidth]{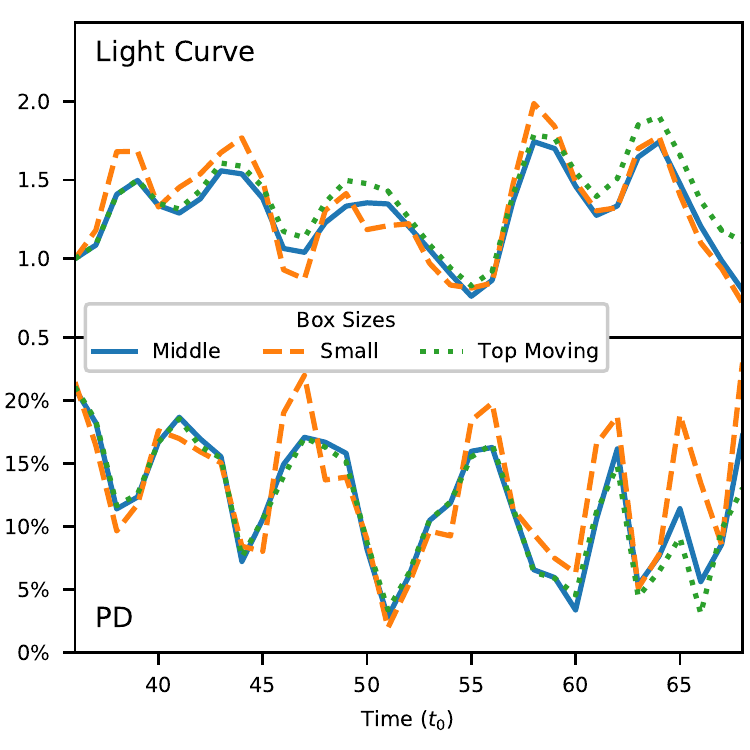}
\caption{light curve (top panel) and PD versus time (bottom panel) for different box sizes; show QPO is not caused by our selection of dissipation site. }
\label{fig:diff_box}
\end{figure}

As seen in Figure~\ref{fig:box_sizes}, the width of the default tracing box (solid blue) is adequate to capture all jet emission in the transverse direction. It is the height and motion of the box that may affect radiation signatures. Here we pick a second smaller box (dashed orange in Figure~\ref{fig:diff_box}), about half of the default box, which centers at the same location and moves up at the same speed as the default tracing box. This is to examine if QPO signatures may be affected when the emission region is only captured partially. Additionally, we pick a third box (dotted red in Figure~\ref{fig:diff_box}) whose bottom edge is fixed at a sufficiently low position to capture all jet radiation, but the top edge is moving up with the default tracing box. This is to examine how QPOs may change with a larger tracing box that covers more than the strongest kinked region. Moreover, since the third box is not moving but has fixed bottom edge, we do not have to worry if some emission may drop out of the tracing box due to its motion. Aforementioned, we cannot use a stationary large box to study radiation signatures, because our jet naturally grows from the central engine, which leads to some shocks at its front. Such energy release is clearly not from kink instabilities, which should not be included in our radiation analyses.

The resulting light curves and PD are plotted in Figure~\ref{fig:box_sizes}. Comparing the radiation signatures from the above two tracing boxes with the default case, we find minor differences and QPOs are preserved. Clearly, both QPOs in light curve and PD are generically from kink instabilities regardless of different tracing boxes. Additionally, the anti-correlated PD with light curves and average PD at $\sim 10\%$ stay the same. Therefore, we conclude that the tracing box does not affect overall radiation and QPO signatures.

\subsection{Implications for Observations}

QPO signatures have been reported in several blazars in various observational bands \citep[e.g.,][]{King13, Ackermann_2015,Graham15}. These signatures have generally been interpreted as quasi-periodic physical processes in the central engine, such as the presence of binary supermassive black holes or the precession of the jet. These interpretations usually suggest that QPOs should appear to be a persistent feature in the source. In observation, however, some potential QPO signatures reported in literature only exhibit several periods. These QPOs can have a wide range of periods, from days/weeks \citep{Hayashida98,zhou_2018,Pihajoki13,Lainela99} up to years \citep{Ackermann_2015, Sandrinelli_2017, Graham15,King13,Raiteri11}. 

In this paper, we show that kink instabilities in blazar jets may lead to QPO signatures. We observe five periods of QPOs in our simulation. Based on our analysis, the period of QPOs in the observer's frame can be estimated as
\begin{equation}
P_{obs} = \frac{R_{\rm KI}}{v_{\rm tr}\delta}~~.
\end{equation}
Here $R_{\rm KI}$ is on the order of the size of the emission region in the comoving frame of the jet. For typical blazar parameters with emission region radius at $10^{16}-10^{17}~\rm{cm}$ and bulk Lorentz factor $\Gamma\sim 10$, the typical period ranges from weeks to months in the observer's frame. Interestingly, QPOs in blazars on such time scales have been reported in several sources on various observational bands \citep{Hayashida98,Lainela99,Rani_2009, zhou_2018}. Notice that, however, kink instabilities may not be a persistent physical process in relativistic jets. In reality, the energy injection into the kinked jet can vary in time, which will strongly affect development of kink instabilities. Therefore, it is unclear whether QPO signatures driven by kink instabilities can be persistent or only last for several periods.

Additional diagnostics for kink-driven QPOs are polarization signatures. The PD generally stays at a relatively low level during kink instabilities. In particular, the PD shows an anti-correlation with the light curve, where it drops during flares. Such a feature has been hinted in several observations \citep{Gaur_2014}. In addition, we find evidence that the PA may also undergo major rotation during kink-driven QPOs. Interestingly, RoboPol team has found that large PA swings are typically accompanied by flares, where the PD can drop during the PA swing \citep{Blinov16,Blinov18}. Based on our simulations, it is possible that the anti-correlated PD with QPO light curves may also appear quasi-periodic. While additional simulations are necessary to further diagnose the polarization variability during kink-driven QPOs, we suggest that future observations of blazar QPO signatures should also include optical polarimetry, so as to diagnose whether the QPOs are driven by kink instabilities.

In addition to blazars, QPOs are also observed in light curves of tidal disruption events (TDE) \citep{Reis_2012, Pasham_2019}. Pevious studies has shown that TDEs could also launch jets \citep{Bloom_2011, Giannios11, Mimica_2016}, which means observed QPOs might come from the kinked region of fresh-grown magnetized jets.  Unlike blazar jets, where the jet might be long standing, TED jets are freshly grown from the central engine such that they are more similar to our RMHD setup. The size of jetted TDE is poorly known, but by taking 200 s as a typical period and a typical of bulk $\Gamma \sim 10$ for TDE jets, we can constrain the size of the TDE jet emission region to be $\sim 1$AU,  which corresponds to $\sim 50$ gravitational radii for a putative $10^6M_\odot$ black hole. Similar to blazar jets, PD monitoring could be good evidence of kink instability's role in QPOs from jetted TDEs. 

\bibliographystyle{mnras} 
\bibliography{jet.bib} 

\appendix
\section*{Acknowledgements} 

We thank Rodolfo Barniol Duran for sharing the RMHD simulation results. The authors acknowledge support from the NASA ATP NNX17AG21G, Fermi Guest Investigator program cycle 12, grant no 80NSSC19K1506, the NSF AST-1910451 and the NSF AST-1816136 grants. HZ acknowledges support from Fermi Guest Investigator program Cycle 11, grant No. 80NSSC18K1723. Simulations are carried out on Purdue Rosen Center for Advanced Computing (RCAC) clusters.

\label{lastpage}
\end{document}